\documentclass{cs19proc}

\usepackage{graphicx}
\usepackage{natbib}
\usepackage{txfonts}
\usepackage{color}
\usepackage{url}

\editors{G.~A. Feiden}
\publisher{Zenodo}
\conference{The 19th Cambridge Workshop on Cool Stars, Stellar Systems, and the Sun}
\conferencedate{2016}

\title{The solar--stellar connection: Magnetic activity of seismic solar analogs}
\author{D. Salabert$^{1}$, 
        R.~A. Garc\'ia$^{1}$,
        P.~G. Beck$^{1}$,
        C. R\'egulo$^{2,3}$,
        J. Ballot$^{4,5}$,
        O.~L. Creevey$^{6}$,
        R. Egeland$^{7,8}$,
        J.-D.~do\,Nascimento Jr.$^{9,10}$,
        F. P\'erez Hern\'andez$^{2,3}$,
        L. Bigot$^{6}$,
	S. Mathur$^{11}$,
        T.~S. Metcalfe$^{11,12}$,
        E.~Corsaro$^{1,2,3}$,		
	P.~L. Pall\'e$^{2,3}$
        }

\affiliation{$^{1}$ Laboratoire AIM, CEA/DRF-CNRS, Universit\'e Paris 7 Diderot, IRFU/SAp, Centre de Saclay, 91191 Gif-sur-Yvette, France\\
$^{2}$ Instituto de Astrof\'isica de Canarias,  E-38200 La Laguna, Tenerife, Spain\\
$^{3}$ Departamento de Astrof\'isica, Universidad de La Laguna, E-38205 La Laguna, Tenerife, Spain\\
$^{4}$ CNRS, Institut de Recherche en Astrophysique et Plan\'etologie, 14 avenue Edouard Belin, 31400 Toulouse, France\\
$^{5}$ Universit\'e de Toulouse, UPS-OMP, IRAP 31400, Toulouse, France\\
$^{6}$ Laboratoire Lagrange, UMR7293, Universit\'e de la C\^ote d'Azur, CNRS, Observatoire de la C\^ote d'Azur, Nice, France\\
$^{7}$ High Altitude Observatory, National Center for Atmospheric Research, PO Box 3000, Boulder, CO 80307-3000, USA\\
$^{8}$ Department of Physics, Montana State University, Bozeman, MT 59717-3840, USA\\
$^{9}$ Universidade Federal do Rio Grande do Norte, UFRN, Dep. de F\'{\i}sica, DFTE, CP1641, 59072-970, Natal, RN, Brazil \\
$^{10}$ Harvard-Smithsonian Center for Astrophysics, Cambridge, Massachusetts 02138, USA\\
$^{11}$ Space Science Institute, 4750 Walnut street Suite\#205, Boulder, CO 80301, USA\\
$^{12}$ Visiting Scientist, National Solar Observatory, 3665 Discovery Dr., Boulder, CO 80303, USA

}

\shorttitle{Magnetic activity of seismic solar analogs}
\shortauthors{D. Salabert et al.}

\abs{Finding solar-analog stars with fundamental properties as close as possible to the Sun and studying the characteristics of their surface magnetic activity is a very promising way to understand the solar variability and its associated dynamo process. However, the identification of solar-analog stars depends on the accuracy of the estimated stellar parameters. Thanks to the photometric CoROT and {\it Kepler} space missions, the addition of asteroseismic data was proven to provide the most accurate fundamental properties that can be derived from stellar modeling today. Here, we present our latest results on the solar--stellar connection by studying 18 solar analogs that we identified among the {\it Kepler} seismic sample \citep{salabert16a}. We measured their magnetic activity properties using the observations collected by the {\it Kepler} satellite and the ground-based, high-resolution \textsc{Hermes} spectrograph. The photospheric ($S_\mathrm{ph}$) and chromospheric ($\mathcal{S}$) magnetic activity proxies of these seismic solar analogs are compared in relation to the solar activity. We show that the activity of the Sun is comparable to the activity of the seismic solar analogs, within the maximum-to-minimum temporal variations of the 11-year solar activity cycle. Furthermore, we report on the discovery of temporal variability in the acoustic frequencies of the young (1 Gyr-old) solar analog KIC\,10644253 with a modulation of about 1.5 years, which agrees with the derived photospheric activity $S_\mathrm{ph}$ \citep{salabert16b}. It could be the signature of the short-period modulation, or quasi-biennal oscillation, of its magnetic activity as observed in the Sun and in the 1-Gyr-old solar analog HD\,30495. In addition, the lithium abundance and the chromospheric activity estimated from \textsc{Hermes} confirms that KIC\,10644253 is a young and more active star than the Sun.
}

\begin{document}

\maketitle

\section{Introduction}
The Mount Wilson Observatory (MWO) monitoring \citep{wilson78,duncan91} of chromospheric activity of main-sequence G and K stars has suggested the existence of two distinct branches of cycling stars, the active and inactive \citep{saar92,soon93}, and that the Sun lies squarely between the two, thus appearing as a peculiar outlier \citep{bohm07}. Today, whether the solar dynamo and the related surface magnetic activity are typical or peculiar still remains an open question \citep{metcalfe16}. Finding solar-analog stars with fundamental properties as close as possible to the Sun and studying the characteristics of their surface magnetic activity is a very promising way to understand solar variability and its associated dynamo \citep{egeland16}. Moreover, the study of the magnetic activity of solar analogs is also important for understanding the evolution of the Sun and its environment in relation to other stars and the habitability of their planets. However, the identification of solar-analog stars depends on the accuracy of the estimated stellar parameters.

The unprecedented quality of the continuous four-year photometric observations collected by the {\it Kepler} satellite \citep{borucki10} allowed the measurements of acoustic oscillations in hundreds of solar-like stars \citep{chaplin14}. Moreover, the length of the {\it Kepler} dataset provides a unique source of information for detecting magnetic activity and the associated temporal variability in the acoustic oscillations. Indeed, it is well established that in the case of the Sun, acoustic (p) oscillation frequencies are sensitive to changes in the surface magnetic activity \citep{wood85}. Moreover, the p-mode frequencies are the only proxy that can reveal inferences on sub-surface changes with activity that are not detectable at the surface by standard proxies \citep[e.g.,][]{salabert09,salabert15,basu12}.

\begin{figure*}[t]
\begin{center} 
\includegraphics[width=0.48\textwidth]{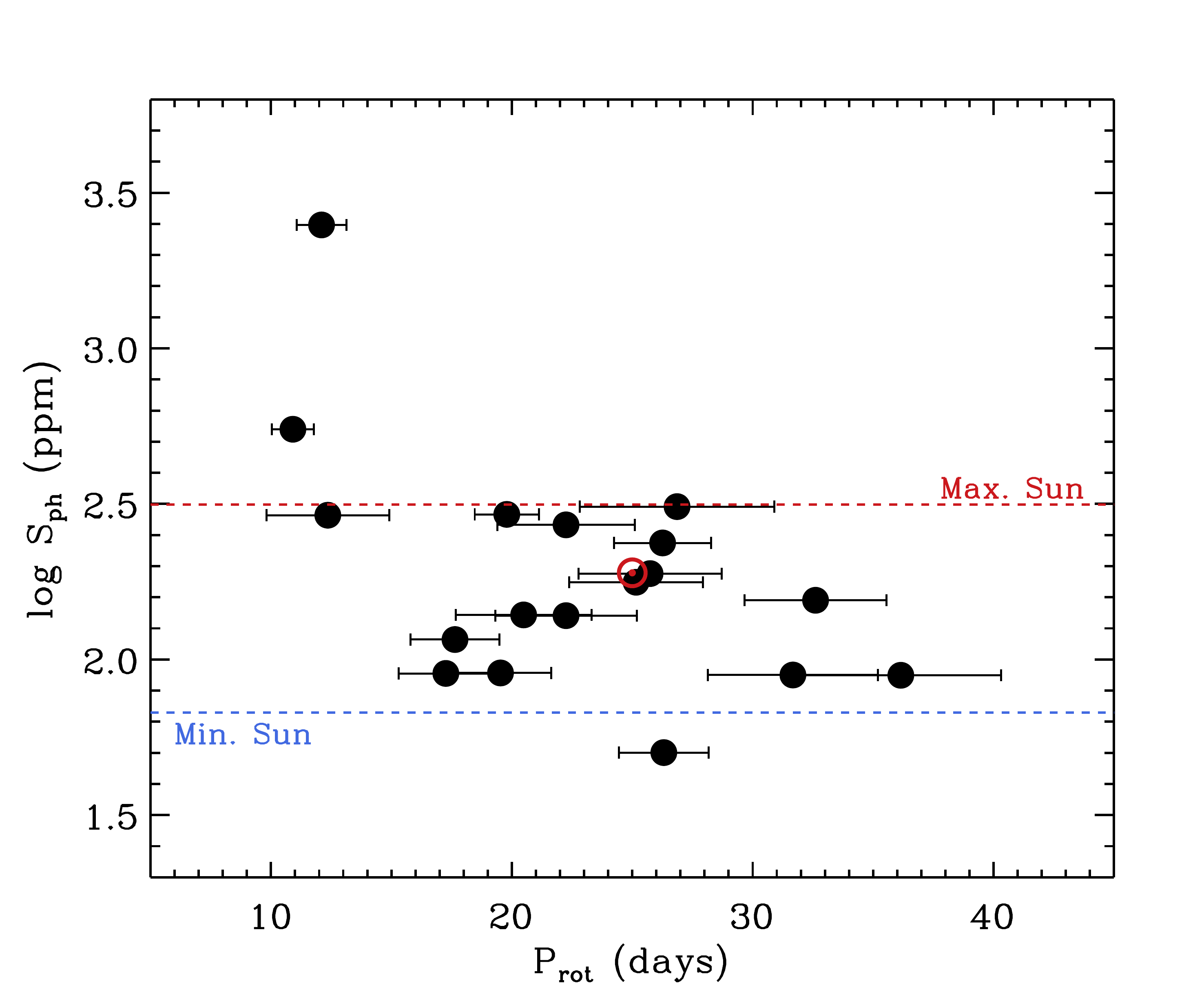}
\includegraphics[width=0.48\textwidth]{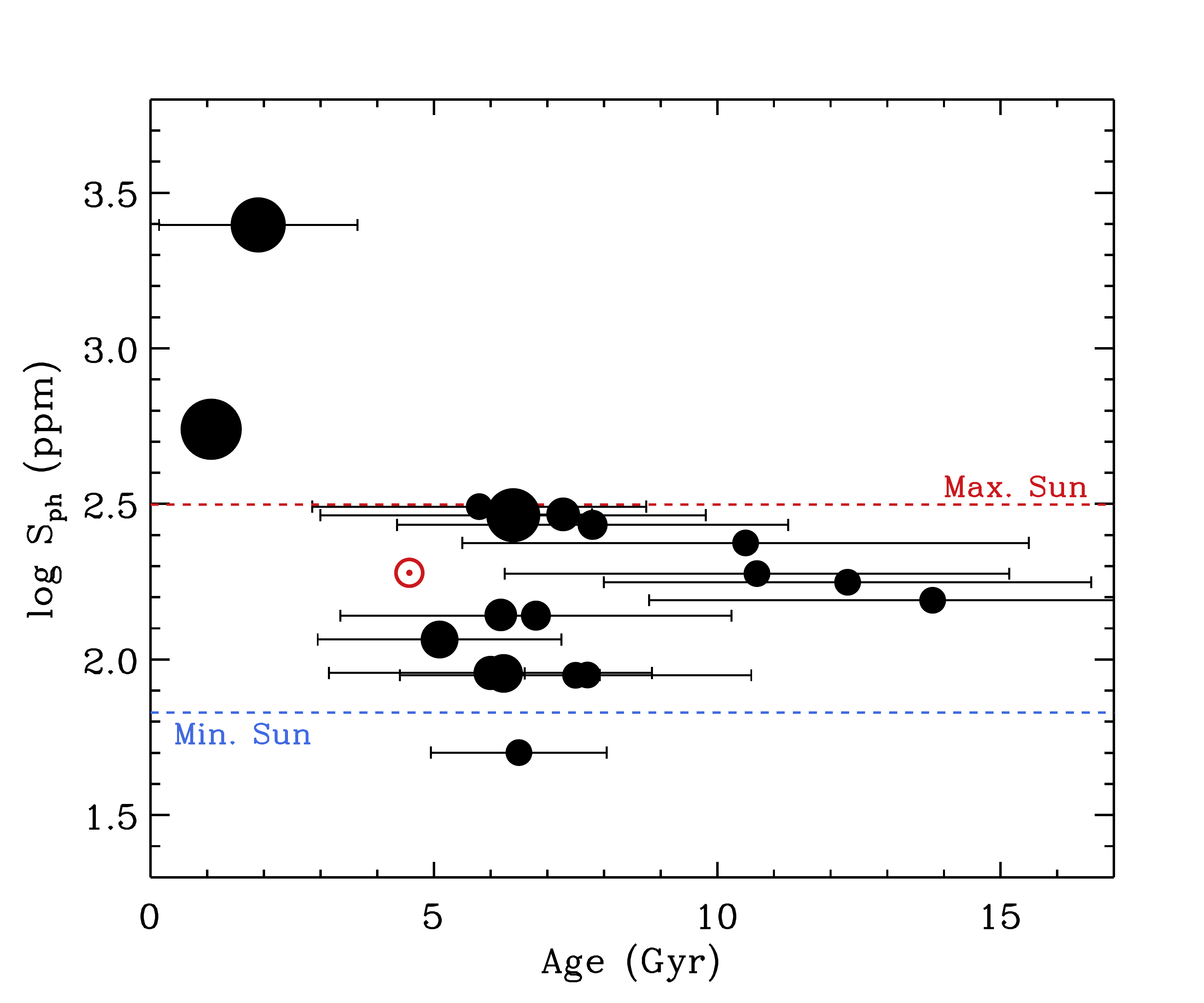}
\end{center}
\caption{{\it Left panel}: Photospheric magnetic activity $S_{\textrm{ph}}$ index as a function of the rotational period $P_\mathrm{rot}$ of the 18 seismic solar analogs identified within the {\it Kepler} sample of solar-like stars. The mean activity level of the Sun calculated from the VIRGO/SPM observations is represented for a rotation of 25\,days with its astronomical symbol (in red), and its mean activity levels at minimum and maximum of the 11-year cycle are represented by the horizontal dashed lines. {\it Right panel}: Same as the left panel but as a function of the seismic ages. The position of the Sun is also indicated for an age of 4.567\,Gyr. The size of the symbols is inversely proportional to the surface rotation $P_\mathrm{rot}$. Adapted from \citet{salabert16a}.} \label{fig:fig1}  
\end{figure*}

\citet{cayrel96} provided a definition of a solar-analog star based on the fundamental parameters, such as mass and effective temperature. Here, we took advantage of the combination of asteroseismology with high-resolution spectroscopy which substantially improves the accuracy of the stellar parameters and reduces their errors \citep{mathur12,chaplin14,metcalfe14,creevey16}. 
Moreover, \citet{mosser09} and \citet{garcia10} show evidences that the magnetic activity inhibits the amplitudes of solar-like oscillations, what is later on confirmed on a large sample by \citet{chaplin11}. In the case of the Sun, the amplitudes of the acoustic modes decrease by about 15\% between solar minimum and maximum for the low-degree modes \citep[see e.g.,][and references therein]{salabert03}, which are the modes detectable in observations of solar-like stars. In consequence, the selection of stars analog to the Sun should include the detection of solar-like oscillations as an additional selection criterium.  This is what we called a seismic solar-analog star, thus extending the \citet{cayrel96}'s definition. Furthermore, we included in the sample only stars with a measured rotation period \citep{garcia14} to ensure the presence of surface magnetic activity.
A total of 18 seismic solar analogs were thus identified \citep{salabert16a} from the photometric {\it Kepler} observations of solar-like oscillators \citep{chaplin14}. 

\section{Magnetic activity of seismic solar analogs}
\subsection{Photospheric magnetic activity}
\label{sec:sph}
Based on the rotation period, $P_\mathrm{rot}$, of a given star, \citet{mathur14b} defined a photospheric proxy of stellar magnetic variability, called $S_\mathrm{ph}$, derived from the analysis of the light-curve fluctuations over sub series of length $5 \times P_\mathrm{rot}$. Its calculation was adapted from the starspot proxy proposed by \citet{garcia10} where they show that the photometric proxy of the F-type HD\,49933 observed by the Convection, Rotation, and planetary Transits \citep[CoRoT,][]{baglin06} satellite is correlated with the measured p-mode frequency shifts and anti-correlated with the mode amplitudes leading them to conclude that the proxy is related to magnetic activity. 
As the variability in the light curves can have different origins and timescales (magnetic activity, convection, oscillations, companion), the rotation period needs to be taken into account. The $S_\mathrm{ph}$ proxy was used to estimate the magnetic activity of several {\it Kepler} targets \citep{mathur14b,garcia14,salabert16a,salabert16b,lund16} and CoRoT \citep{ferreira15} targets, where, for instance, spectral observations used to derive established activity proxies (as the $\mathcal{S}$ and $R'_{\rm HK}$ indices) do not exist and are difficult to obtain for a large number of faint stars. We note however that $S_\textrm{ph}$ represents a lower limit of the photospheric activity as it depends on the inclination angle of the star's rotation axis in respect to the line of sight. This is however assuming that the development of the latitudinal distribution of the starspots is comparable to the one observed for the Sun, that is from mid to low latitudes.
Other photospheric metrics were developed to study the stellar variability in the {\it Kepler} data but they do not use the knowledge of the rotation rate in their definition \citep{basri10,basri11,chaplin11,campante14}. 
Here, the photospheric activity $S_\textrm{ph}$ index of the 18 identified seismic solar analogs was estimated through the analysis of the {\it Kepler} long-cadence observations which were calibrated using the \emph{Kepler} Asteroseismic Data Analysis and Calibration Software \citep[KADACS,][]{garcia11}. 

The left panel of Fig.~\ref{fig:fig1} shows the $S_\textrm{ph}$ of the 18 seismic solar analogs as a function of their surface rotation period $P_\mathrm{rot}$. The mean value of the solar $S_\textrm{ph}$ over cycle~23  calculated from the photometric observations collected by the Variability of Solar Irradiance and Gravity Oscillations  \citep[VIRGO;][]{frohlich95} instrument onboard the Solar and Heliospheric Observatory \citep[SoHO;][]{domingo95} spacecraft is also indicated, as well as the corresponding values at minimum and maximum of activity. We note that the VIRGO instrument is composed of three Sun photometers (SPM). The photospheric activity of the identified 18 seismic solar analogs is comparable to the Sun, within the range of activity covered over a solar cycle. The right panel of Fig.~\ref{fig:fig1} shows for the same set of 18 stars the measured photospheric activity as a function of their seismic ages \citep{mathur12,chaplin14,metcalfe14} compared to the Sun. We note the large differences in the uncertainties of the age estimates depending on the applied method to derive them \citep[see][]{metcalfe14}. It is also worth noticing that stars between 2 and 5\,Gyr-old are missing in our sample.
Nevertheless, the position of the Sun indicates that its photospheric activity is compatible with older solar analogs. Although the {\it Kepler} seismic sample contains only few stars younger than the Sun, our results show that the two youngest seismic solar analogs in our sample below 2\,Gyr-old are actually the most active, as well as being the fastest rotating stars. Furthermore, the photospheric activity of stars older than the Sun seems to not evolve much with age.
\citet{pace13} studied the chromospheric activity of field dwarf stars as an age indicator, and shows that it stops decaying for stars older than 2\,Gyr-old. 
The photospheric activity of the solar analogs analyzed here is observed to be comparable to the Sun after 2\,Gyr-old within the minimum-to-maximum range of the solar cycle, given the uncertainties on the age estimates and the limited size of our sample. This is also assuming that our sample is not too biased towards stars observed during periods of low activity of their magnetic cycles, or that they are not in an extended cycle. Nevertheless, observations of additional solar analogs younger than the Sun are needed to fill the range of ages below 5\,Gyr-old.

\begin{figure}[tbp]
\begin{center} 
\includegraphics[width=0.49\textwidth]{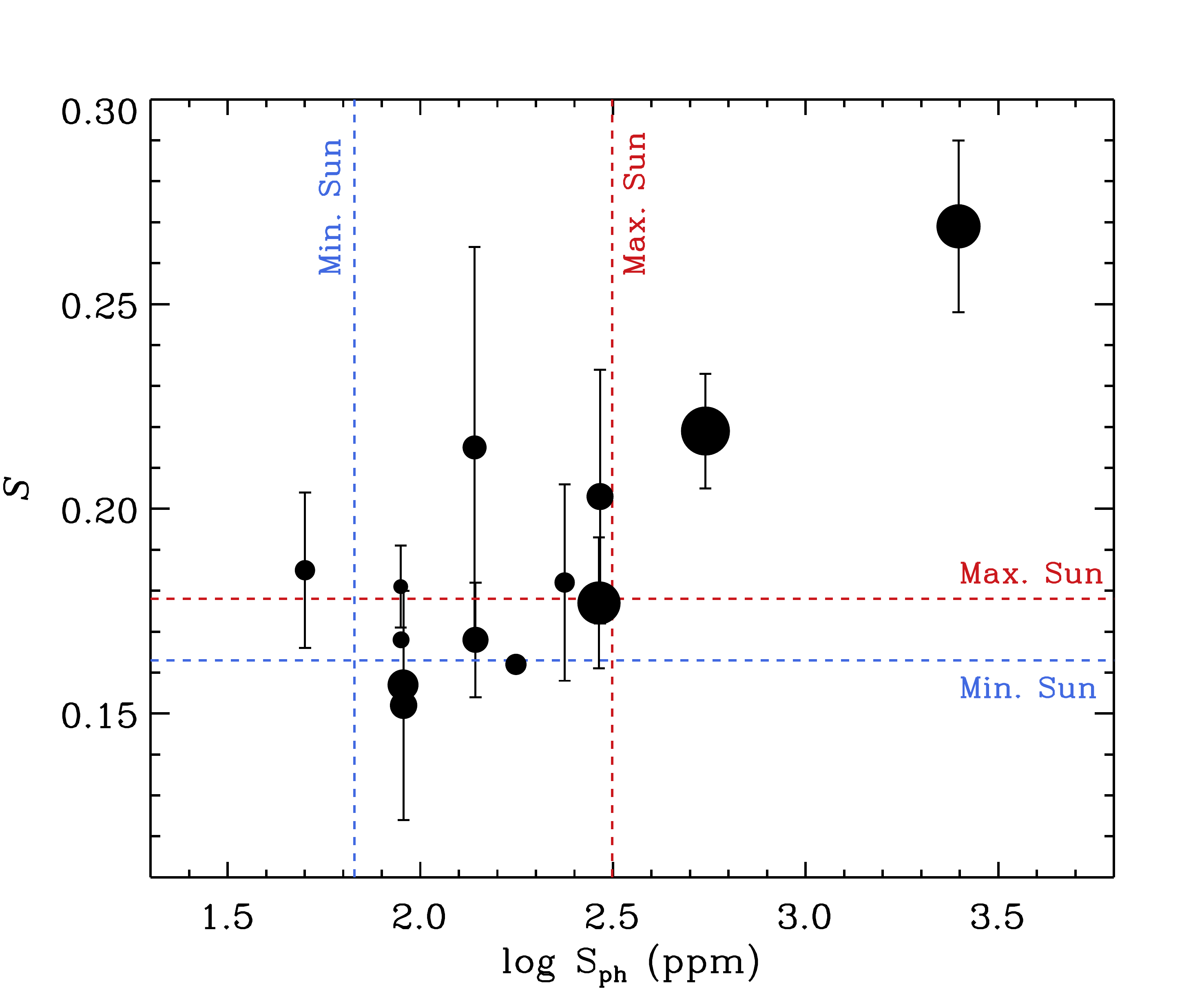}
\end{center}
\caption{Chromospheric $\mathcal{S}$~index derived from \textsc{Hermes} observations and calibrated into the MWO system as a function of the photospheric $S_\textrm{ph}$ index. The corresponding activity levels of the Sun at minimum and maximum
of its 11-year magnetic cycle are represented by the horizontal
and vertical dashed lines. The symbol size is inversely proportional to the rotation period $P_\mathrm{rot}$. Adapted from \citet{salabert16a}.} \label{fig:fig2}  
\end{figure}

\subsection{Chromospheric magnetic activity}
In addition, the chromospheric activity $\mathcal{S}$ index of these 18 solar analogs was estimated from spectroscopic observations collected with the \textsc{Hermes} spectrograph \citep{raskin11} mounted on the 1.2-m \textsc{Mercator} telescope at the Observatorio del Roque de los Muchachos (La Palma, Canary Islands, Spain). A detailed description of the data processing of the \textsc{Hermes} observations can be found in \citet{beck16}. The $\mathcal{S}$ index measures the strength of the plasma emission in the cores of the Ca\textsc{ii}\,H\&K lines in the near ultra violet \citep{wilson78}. The result is dependent on the instrumental resolution and on the spectral type of the star. However, as the selected stars were chosen for all having comparable stellar properties to the Sun, the estimated values of the $\mathcal{S}$~index can be thus safely compared between each others.

The comparison between the photospheric $S_\mathrm{ph}$ and the chromospheric $\mathcal{S}$ magnetic activity proxies is shown on Fig.~\ref{fig:fig2} for a subset of 13 stars with a signal-to-noise ratio (S/N) in the near ultra violet S/N(Ca)\,$>$\,15 \citep[see details in][]{salabert16a}. We note that the $\mathcal{S}$ index  was scaled to the MWO system following the \textsc{Hermes} instrumental scaling factor derived by \citet{beck16}. The mean values at minimum and maximum along the solar cycle of the photospheric (see Section~\ref{sec:sph}) and chromospheric \citep[Egeland et al., submitted;][]{hall04} magnetic activity levels are also represented. The resulting activity box corresponds to the range of change in solar activity along the 11-year magnetic cycle. Although the sample of stars is small, the $S_\mathrm{ph}$ and $\mathcal{S}$ indices are observed to be complementary, within the errors. We note also that both proxies were not estimated from contemporaneous {\it Kepler} and \textsc{Hermes} observations, introducing a dispersion partly related to possible temporal variations in the stellar activity. Nevertheless, it confirms that $S_\mathrm{ph}$ can complement the classical $\mathcal{S}$~index for activity studies. 
Furthermore, the level of activity of solar analogs observed with the Mount Wilson program \citep{baliunas96,donahue96} is in agreement with our {\it Kepler} sample, giving additional confidence in our determination of the activity.

\begin{figure}[tbp]
\begin{center} 
\includegraphics[width=0.48\textwidth]{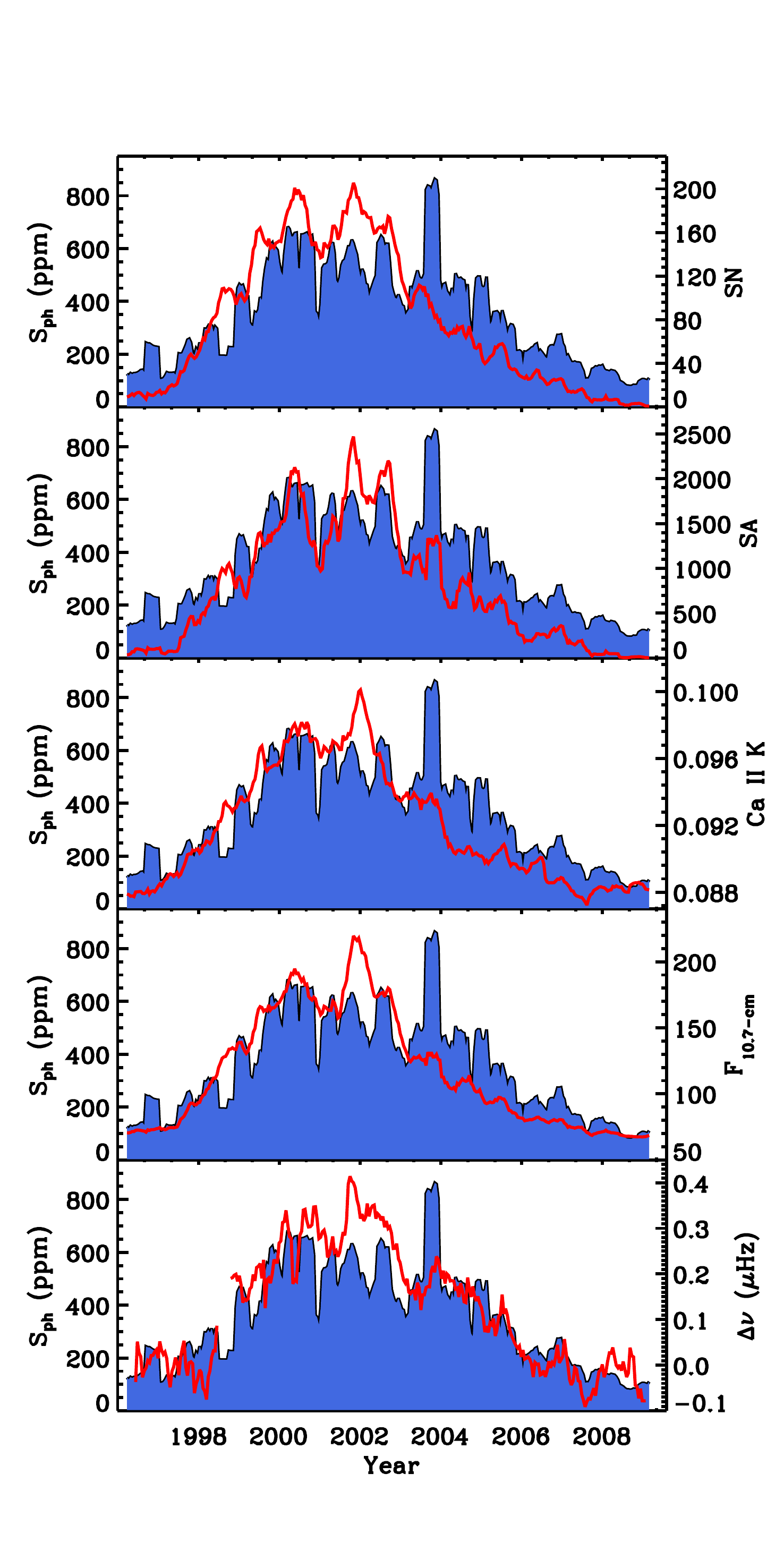}
\end{center}
\caption{Photospheric activity proxy $S_{\textrm{ph}}$ measured from the photometric VIRGO/SPM observations (blue filled) as a function of time and compared to some standard indices of solar activity (red solid lines). {\it From top to bottom}: the right-hand y axis corresponds to the sunspot number (SN); the sunspot area (SA); the Ca\textsc{ii}\,K-line emission index in $\AA$; the 10.7-cm radio flux (\textrm{F$_\textrm{10.7-cm}$}) in 10$^{-22}$~s$^{-1}$~m$^{-2}$~Hz$^{-1}$;  and the frequency shifts $\Delta\nu$ of the acoustic low-degree oscillations in $\mu$Hz.} \label{fig:fig3}  
\end{figure}

\section{Comparison with solar magnetic activity}
The photospheric activity proxy $S_\textrm{ph}$ of the Sun was calculated for sub series of $5\,\times\,P_\mathrm{rot}\,=\,125$~days (with $P_\mathrm{rot}\,=\,25$~days) over 4720~days of the VIRGO/SPM observations covering the entire solar cycle~23 (1996\,--\,2008). 
The resulting temporal evolution of $S_\mathrm{ph}$ is compared in Fig.~\ref{fig:fig3} to the some standard solar proxies which were averaged over the same sub series: (1) the total sunspot number\footnote{Source:~WDC-SILSO, Royal Observatory of Belgium at \url{http://www.sidc.be/silso/datafiles}}; (2) the total sunspot area\footnote{Source:~\url{http://solarscience.msfc.nasa.gov/greenwch.shtml}}; (3) the chromospheric  Ca\textsc{ii}\,K-line emission\footnote{Source:~Sacramento Peak Observatory at \url{http://nsosp.nso.edu/}}; (4) the 10.7-cm radio flux\footnote{Source:~National Geophysical Data Center at \url{http://www.ngdc.noaa.gov/stp/solar/solardataservices.html}}. We also compared the $S_\mathrm{ph}$ to the temporal variations of the acoustic low-degree oscillation frequencies extracted over the same sub series of VIRGO/SPM observations following the methodology described in \citet{salabert15}. Figure~\ref{fig:fig3} clearly demonstrates that the $S_\textrm{ph}$ index is a good photospheric proxy for solar activity.

\begin{figure}[t]
\begin{center} 
\includegraphics[width=0.49\textwidth]{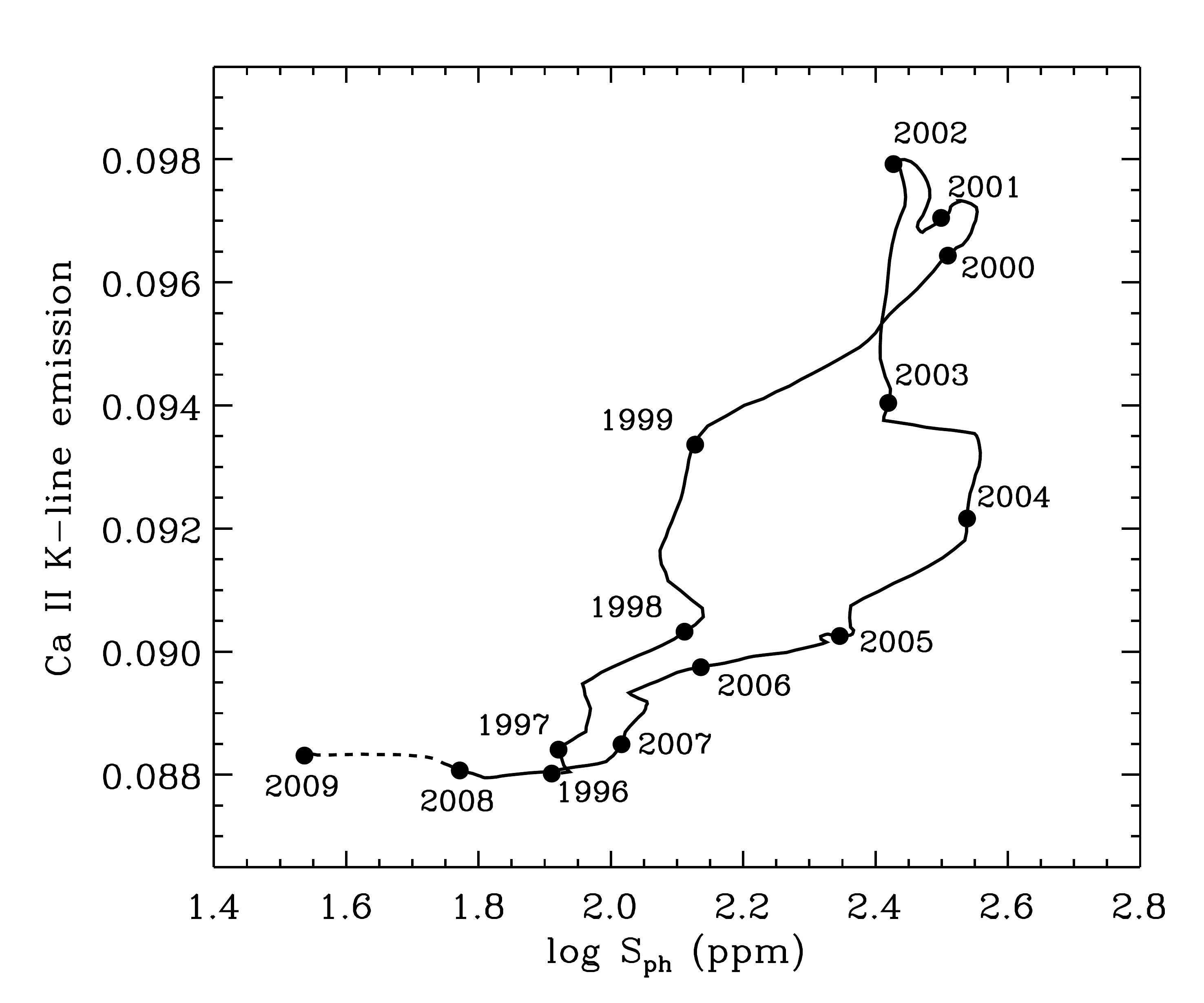}
\end{center}
\caption{Solar Ca\textsc{ii}\,K-line emission as a function of the photospheric magnetic proxy $S_\mathrm{ph}$ of the Sun during the solar cycle 23 (1996--2008) and smoothed over one year. The first year of cycle 24 is also represented in dashed line. The start of each year is indicated. Adapted from \citet{salabert16a}.} 
\label{fig:fig4}  
\end{figure}

Furthermore, to better understand the relation between the photospheric and chromospheric magnetic activity of the solar analogs represented in Fig.~\ref{fig:fig2}, we directly compared the temporal variations of the photospheric $S_\mathrm{ph}$ index of the Sun to the Ca\textsc{ii}\,K-line emission index (Fig.~\ref{fig:fig4}). As in Fig.~\ref{fig:fig3}, the period covered corresponds to the solar cycle~23. For illustrative purpose, the data points in Fig.~\ref{fig:fig4} were smoothed over a period of one year. The start of each year is also indicated. 
The relationship between the two proxies varies between the rising and falling phases of the solar cycle following an hysteresis pattern.
Such hysteresis is observed in a wide range of solar observations between photospheric and chromospheric activity proxies, as well as with the p-mode frequency shifts \citep[e.g.,][]{bach94,chano98,tripathy01,ozguc12}. Finally, Fig.~\ref{fig:fig4} supports the complementarity between the photospheric $S_\mathrm{ph}$ and chromospheric $\mathcal{S}$ proxies observed for the solar analogs in Fig.~\ref{fig:fig2}.

\begin{figure*}[tbp]
\begin{center} 
\includegraphics[width=0.70\textwidth]{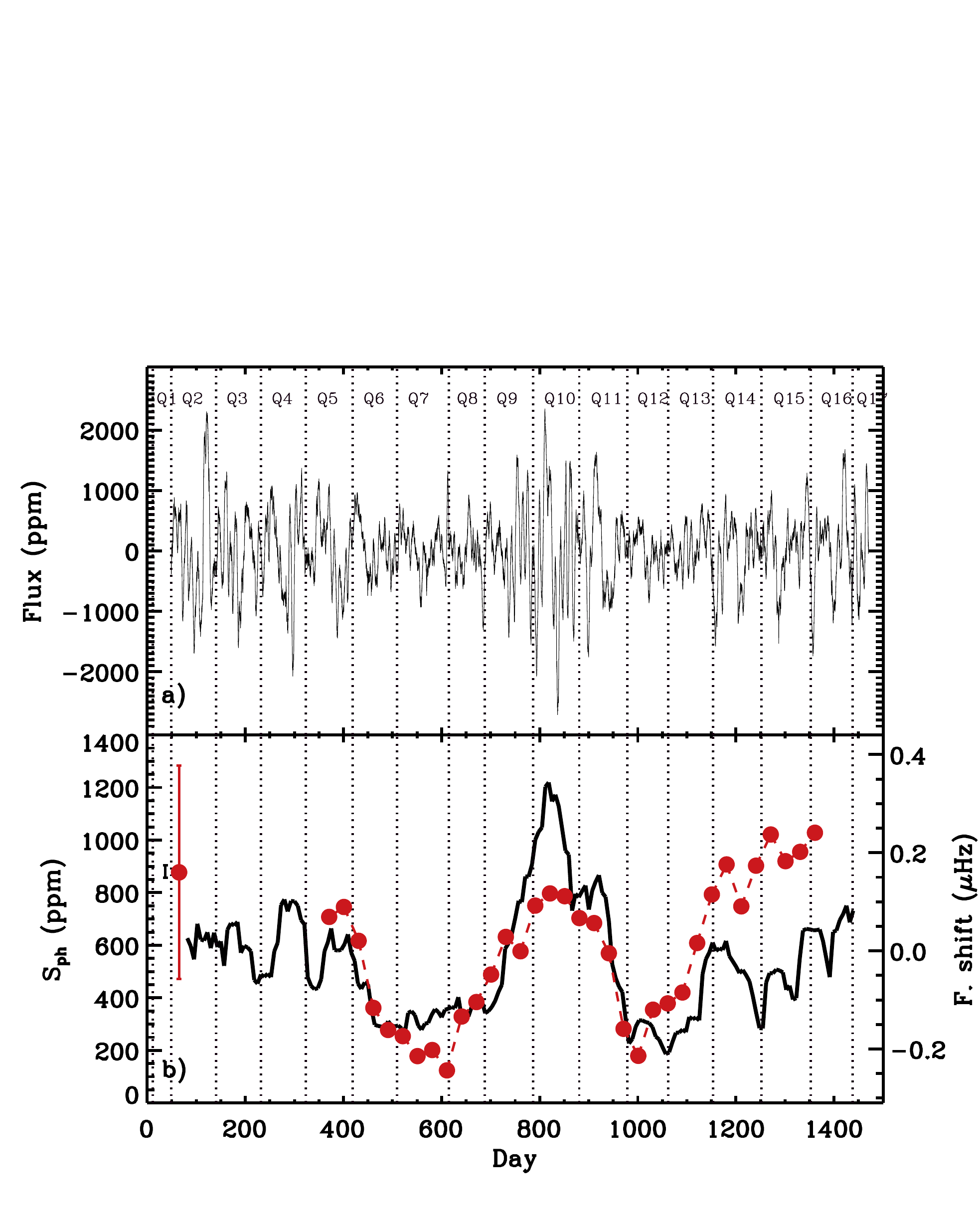}
\end{center}
\caption{\label{fig:fig5} {\it Top panel}: Photometric long-cadence observations of KIC\,10644253 collected over 1411 days by the {\it Kepler} satellite as a function of time. 
{\it Bottom panel}: Photospheric activity $S_{\textrm{ph}}$ index (black) of KIC\,10644253 as a function of time compared to the frequency shifts obtained from the cross-correlation analysis (red circles). The frequency shifts were extracted from the continuous short-cadence observations from Q5 to Q17. The associated mean uncertainties are illustrated in the upper left-hand corner using the same color code. In the two panels, the vertical dotted lines represent the observational length of each {\it Kepler} quarter from Q1 to Q17. Adapted from \citet{salabert16b}.}
\end{figure*} 

\section{Magnetic variability in the young solar analog KIC\,10644253}
With a rotation period of $10.91\,\pm0.87$~days \citep{garcia14}, the solar analog KIC\,10644253 (BD+47\,2683, $V=9.26$) is the youngest solar-like pulsating star observed by {\it Kepler} with an age of $1.07\,\pm\,0.25$\,Gyr \citep{metcalfe14} and one of the most active \citep{garcia14}. KIC\,10644253 is thus an excellent candidate for investigating the magnetic activity of a young Sun with asteroseismic data. 
In addition to the Sun, temporal variations of p-mode frequencies related to magnetic activity were so far observed in only three stars: the F-type stars HD\,49933 \citep{garcia10} and KIC\,3733735 \citep{regulo16}, and the solar-analog G-type KIC\,10644253 \citep{salabert16b}. 

To study the temporal variations of the low-degree, p-mode oscillation frequencies observed in KIC\,10644253, the {\it Kepler}  short-cadence dataset was split into contiguous 180-day-long sub series. The associated power spectra were analyzed using both peak-fitting \citep{ballot11,salabert11} and cross-correlation \citep{regulo16} independent methods and the corresponding frequency shifts $\Delta \nu$ extracted \citep{salabert16b}. In addition, the light curve was analyzed to estimate the $S_\mathrm{ph}$ over sub series of $5\,\times\, P_\mathrm{rot}\,=\,54.55$\, days.
Figure~\ref{fig:fig5} shows that both the photospheric $S_\mathrm{ph}$ and the frequency shifts $\Delta\nu$ present the signature of magnetic activity with a significant temporal variability. A modulation of about 1.5\,years  is measured in both observables of about 900\,ppm for $S_\mathrm{ph}$ and 0.5\,$\mu$Hz for the frequency shifts. It could be the signature of the short-period modulation, or quasi-biennal oscillation, of its magnetic activity as observed in the Sun \citep[see, e.g.,][]{fletcher10}. The variations found in KIC\,10644253 at a rotation period of about 11~days is analogous to what is found by \citet{egeland15} from the study of the temporal variations of the Mount Wilson $\mathcal{S}$ index in the solar analog HD30495 falling on the inactive branch \citep{bohm07}. Moreover, the comparison between magnitude and frequency dependence of the
frequency shifts measured for KIC\,10644253 with the ones obtained for the Sun indicates that the same physical mechanisms are involved in the sub-surface layers in both stars. 

The analysis of the spectroscopic observations collected by the \textsc{Hermes} instrument shows that KIC\,10644253 is about 18\% chromospherically more active than the Sun with a $\mathcal{S}$ index of $0.213\pm0.008$. Its high lithium abundance of $2.74\pm0.03$\,dex and its effective temperature of $6006\pm100$\,K  mean that the lithium at the surface has not been depleted yet by internal processes \citep{ramirez12}.  This is thus validating its young age estimated from seismology and in agreement with a rotation of about 11~days from gyrochronology \citep{meibom11,jen16}.  When comparing the measured lithium abundance with its theoretical evolution, Beck et al. (submitted) found a good consensus with the age and mass derived from asteroseismology. Furthermore, among the 18 solar analogs in this sample, KIC\,10644253 has the highest lithium abundance. See also Beck et al. in these proceedings.

\section{Conclusions}
The characteristics of the surface magnetic activity of solar analogs can provide useful new constraints to better understand the variability of the Sun and its underlying dynamo. Here, we analyzed the sample of main-sequence stars observed by the {\it Kepler} satellite for which solar-like oscillations were detected \citep{chaplin14} and rotational periods measured \citep{garcia14}. We identified 18 seismic solar analogs \citep{salabert16a} from the asteroseismic stellar properties found in the literature \citep{mathur12,chaplin14,metcalfe14}. We then studied the properties of the photospheric and chromospheric magnetic activity of these stars in relation of the Sun \citep{salabert16a}. The photospheric activity proxy $S_\mathrm{ph}$ was derived by means of the analysis of the {\it Kepler} observations, while the chromospheric activity proxy $\mathcal{S}$ was measured with follow-up, ground-based  spectroscopic observations collected by \textsc{Hermes} spectrograph.

We showed that the magnetic activity of the Sun is comparable to the activity of the seismic solar analogs studied here, within the maximum-to-minimum activity variations of the Sun during the 11-year cycle. As expected, the youngest and fastest rotating stars are observed to actually be the most active of our sample. The comparison of the photospheric $S_\mathrm{ph}$ with the well-established chromospheric $\mathcal{S}$ shows that the $S_\mathrm{ph}$ index can be used to provide a suitable magnetic activity proxy. Moreover, the $S_\mathrm{ph}$ can be easily estimated for a large number of stars with known surface rotation observed simultaneously with photometric space missions, while the estimation of the associated $\mathcal{S}$ index would be highly difficult to achieve as it would require a lot of time of ground-based telescopes to collect enough spectroscopic data for each individual target. Further, these photometric targets are rather faint for spectroscopy observations.

The properties of the young (1 Gyr-old) solar analog KIC\,10644253 make this star an excellent candidate for investigating the solar--stellar connection by studying the temporal variability of the magnetic activity of a young Sun with asteroseismic data \citep{salabert16b}. Indeed, we showed the existence of a modulation of about 1.5 years which is measured in both the low-degree, p-mode frequency shifts $\Delta\nu$ and in the photospheric activity $S_\mathrm{ph}$ index. This modulation could be the signature of the short-period modulation of its magnetic activity as observed in the Sun  \citep[see, e.g.,][]{fletcher10} and in the 1-Gyr-old Mount Wilson solar analog HD\,30495 \citep{egeland15} which falls on the inactive branch \citep{bohm07}. In addition, the lithium abundance and the chromospheric activity estimated from \textsc{Hermes} confirms that KIC\,10644253 is a young and more active star than the Sun.

\section*{Acknowledgments}
{The authors wish to thank the entire {\it Kepler} team, without whom these results would not be possible. Funding for this Discovery mission is provided by NASA's Science Mission Directorate. The ground-based observations are based on spectroscopy made with the Mercator Telescope, operated on the island of La Palma by the Flemish Community, at the Spanish Observatorio del Roque de los Muchachos of the Instituto de Astrof\'isica de Canarias. This work utilizes data from the National Solar Observatory/Sacramento Peak CaII K-line Monitoring Program, managed by the National Solar Observatory, which is operated by the Association of Universities for Research in Astronomy (AURA), Inc. under a cooperative agreement with the National Science Foundation.
The research leading to these results has received funding from the European Community's Seventh Framework Programme ([FP7/2007-2013]) under grant agreement no. 312844 (SPACEINN). The research leading to these results has also been supported by grant AYA2012-39346-C02-02 of the Spanish Secretary of State for R\&D\&i (MINECO). DS and RAG acknowledge the financial support from the CNES GOLF and PLATO grants. PGB acknowledges the ANR (Agence Nationale de la Recherche, France) program IDEE (no. ANR-12-BS05-0008) "Interaction Des Etoiles et des Exoplan\`etes". JDNJr acknowledges support from CNPq PQ 308830/2012-1 CNPq PDE Harvard grant.
RE is supported by the Newkirk Fellowship at the High Altitude Observatory.
SM acknowledges support from the NASA grant NNX12AE17G and NNX15AF13G. TM acknowledges support from the NASA grant NNX15AF13G.
DS acknowledges the Observatoire de la C\^ote d'Azur for support during his stays. This  research  has  made  use  of  the
SIMBAD database, operated at CDS, Strasbourg, France.
}

\bibliographystyle{cs19proc}

\end{document}